\documentclass[conference]{IEEEtran}
\IEEEoverridecommandlockouts
\usepackage{cite}
\usepackage{algorithm}
\usepackage{algpseudocode}
\usepackage{amsmath}
\usepackage{amsfonts}
\usepackage{amsmath,amssymb,amsfonts}
\usepackage{bm}
\usepackage{enumerate}
\usepackage{subfigure}
\usepackage{graphicx}
\usepackage{textcomp}

\def\BibTeX{{\rm B\kern-.05em{\sc i\kern-.025em b}\kern-.08em
    T\kern-.1667em\lower.7ex\hbox{E}\kern-.125emX}}

\begin{document}

%\title{ A  Novel Random Access Scheme for  6G Cellular
%Internet of Things Networks  with mMTC and URLLC
% Co-Existence }
\title{Smart City Enabled by 5G/6G Networks: An Intelligent Hybrid Random Access Scheme}

\author{Huimei Han$^{1}$, Wenchao Zhai$^2$,Jun Zhao$^3$
\vspace{1.5mm}
\\
\fontsize{10}{10}\selectfont\itshape
$^1$College of Information Engineering, Zhejiang University of Technology,
Hangzhou, Zhejiang Province, China\\
$^2$College of Information Engineering, China Jiliang University,
Hangzhou, Zhejiang Province, China\\
%$^3$State Key Lab of Integrated Services Networks, Xidian University, Xi¡¯an,
%710071, P.R. China\\
$^3$School of Computer Science and Engineering, Nanyang Technological University, Singapore\\
%$^3$Laboratory of Signals and Systems of Paris-Saclay University - CNRS, CentraleSup\'{e}lec, Univ Paris Sud, France\\
\vspace{1.5mm}
\\
\fontsize{9}{9}%\selectfont\ttfamily\upshape
$^1$ hmhan1215@zjut.edu.cn, $^2$ zhaiwenchao@cjlu.edu.cn
$^3${junzhao}@ntu.edu.sg \vspace{-2mm}
}

%\title{Smart City Enabled by 5G/6G Networks: An Intelligent Hybrid Random Access Scheme}

%\author{\IEEEauthorblockN{Huimei Han, Jun  Zhao, Wenchao Zhai, Zehui Xiong, Weidang Lu}

%\thanks{
%Huimei Han and  Weidang Lu are with College of Information Engineering, 
 %Zhejiang University of Technology, Hangzhou, Zhejiang, 310032, 
% P.R. China. 
 % Jun Zhao and Zehui Xiong are with School of
 %Computer Science and Engineering, Nanyang Technological University, Singapore. Wenchao Zhai is with the China Jiliang University, 
 %Hangzhou, Zhejiang, 310018, P.R. China.
 % (Emails: \{hmhan1215,\,luweid\}@zjut.edu.cn,  \{junzhao,\,zehui.xiong\}@ntu.edu.sg, zhaiwenchao@cjlu.edu.cn)

%Huimei Han  is with College of Information Engineering, Zhejiang University of Technology, Hangzhou, Zhejiang, 310032, P.R. China, Ying Li is with the State Key Lab of Integrated Services Networks, Xidian University, Xi'an, 710071, P.R. China, Wenchao Zhai is with the China Jiliang University, Hangzhou, Zhejiang, 310018, P.R. China,  and Liping Qian is with the Zhejiang University of Technology, Hangzhou, Zhejiang, 310032,  P.R. China  (Email: hmhan1215@zjut.edu.cn, yli@mail.xidian.edu.cn, zhaiwenchao@cjlu.edu.cn, lpqian@zjut.edu.cn)

%}}

\maketitle

\begin{abstract}
 The Internet of Things (IoT) is the enabler for smart city to achieve the envision of the
  ``Internet of Everything'' by intelligently connecting devices without human interventions.
The explosive growth of IoT devices makes the amount of business data generated by machine-type communications (MTC) account
for a great proportion in all communication services.  The fifth-generation (5G) specification for cellular networks defines two types of application scenarios for MTC:
 One is massive machine type communications (mMTC) requiring  massive connections, while the other  is ultra-reliable low latency communications
 (URLLC) requiring high reliability and low latency communications. 6G, as the next generation beyond 5G, will have even stronger scales of mMTC and URLLC. mMTC and URLLC will co-exist in MTC networks for 5G/6G-enabled smart city. To enable massive and reliable LLC access  to  such  heterogeneous  MTC  networks where mMTC and URLLC  co-exist,
 in this article,
 we introduce the network  architecture of heterogeneous MTC networks, and   propose an intelligent hybrid random access scheme 
 for 5G/6G-enabled smart city.
 Numerical results show  that, compared to the  benchmark schemes, the proposed scheme significantly improves the successful access probability, and satisfies the  diverse quality of  services requirements of URLLC and mMTC devices.

\end{abstract}

\begin{IEEEkeywords}
Internet of Things, Machine-Type Communications, Random Access,
 Smart City.    
\end{IEEEkeywords}

% \fi
\IEEEpeerreviewmaketitle

\section{Introduction}

 With the development of technology and urbanization, smart cities have  become the  development trend of cities and received much attention in recent years.
 A smart city aims to  fully realize intelligence in various social services, such as residents' lives,
  government services, security, and education,
 which makes the city on demand. To realize the smart city, technologies such as the Internet of Things (IoT), cloud computing and  mobile technology,
  are essential. Among these technologies, IoT, as the foundation of the smart city, is attracting increasing attention
 from both industry and academia  due to  its promising applications of  intelligently connecting devices without human interventions~\cite{IOTSC}.
 %The world's second largest market research predicts that the smart city IoT  will be expected to worth USD 219.6 billion by 2023.

According to Gartner prediction,  more than 30 billion IoT devices will connect to  the network by the year 2025~\cite{predict}. Thus,
 the amount of business data generated by machine-type communications (MTC) accounts for a great proportion in all communication services~\cite{IOTC}.
Due to the extensive coverage, security, reliability and flexibility of communication of  the fifth-generation (5G) communication networks, 
5G networks have become
the main force to support and promote  MTC services. The 5G specification has defined two  application scenarios for MTC:
 One is massive machine type communications (mMTC), which aims to provide massive connections; The other  is ultra-reliable low 
 latency communications (URLLC),  which aims to provide high reliability and low latency communications.  mMTC and URLLC  will co-exist
 in MTC network for smart city. For example, driverless and remote surgery require reliable and  
 low latency transmission, while environmental monitoring and smart agriculture need massive access. Recently, 6G cellular communications, as the next generation of  beyond 5G, have been speculated~\cite{MTC}. It is envisioned~\cite{MTC} that in 6G, mMTC and URLLC will still exist, despite at even stronger scales; i.e., compared with 5G, 6G will have even more massive MTC and even more reliable LLC.

%  The network 
%  %by proper resource management, IoT networks 
%  should give priority to  reliable transmissions of driverless and remote surgery.
 %Environmental monitoring, smart agriculture Driverless and remote surgery

%with different quality of service (QoS)
%The connection-oriented random access procedure utilized in the long term evolution (LTE) network aims to provide 
%services for human communications,
Random access is the first  and critical step of communication between the BS  and devices.
% which is a vital part in the communication systems.
%
 The random access mechanism in the traditional cellular  communication system
 is mainly designed for human-to-human (H2H) communication with large data packet transmission, few connections, and low energy consumption requirements.
 In order to support 5G MTC communications  with small data packet transmission, sporadic transmission, and diversified Quality of Service (QoS)
 requirements,
  it is necessary to design an effective random access scheme. Recently, artificial intelligence (AI)-based
approaches have been used to support MTC communications~\cite{ML1,ML3}.  A reinforcement learning based
 random access scheme was proposed to learn the optimal access class barring (ACB) factor under the  different number of active devices~\cite{ML1}.
 Gui~\textit{et~al.}~\cite{ML2} proposed  a long short-term memory (LSTM)-based random access  scheme where LSTM  deep learning model is utilized
 to  allocate power to each device.
%N. Ye~\textit{et~al.}~established an end-to-end NORA network model where deep variational autoencoder is utilized to realize the  uplink message decoding
%and active UE detection~\cite{ML3}.
However, the above works only studied the case of all devices having the same priority, which are not suitable for heterogeneous
MTC network in smart city where URLLC and mMTC devices co-exist~\cite{MTC}.  To enable massive and reliable LLC access  to  such  heterogeneous  MTC  networks,
 we introduce the   architecture of heterogeneous MTC networks, and propose
  an intelligent hybrid random access (IHRA) scheme for 5G-enabled smart city. Our scheme also applies to  6G-enabled smart city, where mMTC and URLLC will still exist, despite at even stronger scales. For simplicity, we focus on 5G in the rest of the paper. 
  %This scheme predicts the number of  active URLLC devices by employing a proposed attention-based LSTM prediction model thereby
   %determining  the parameters of the multi-user detection algorithm dynamically. Thus, URLLC devices
   %access the network via  a two-step contention-free  access procedure, to  meet latency and reliability access requirements;
   % mMTC devices access the network via a contention-based TA-aided access mechanism to meet  massive access requirement.
 %Numerical results show  that,
  %   compared to the benchmark schemes, the LSTMH-RA scheme significantly improves the successful access probability,
   %   and satisfies the  diverse QoS requirements of URLLC and mMTC devices.
 %To enable  accesses of such heterogeneous MTC network, in this article,
% we introduce the network  architecture of heterogeneous MTC network, and then propose an intelligent hybrid random access scheme for 5G-enabled smart city.
%An attention-based  long short-term memory (LSTM) prediction \mbox{model} is introduced to predict the number of active URLLC devices.
The features and {main contributions} of this paper are summarized as follows.
\begin{itemize}
   \item We propose an IHRA scheme for 5G-enabled smart city. More specifically,
 to  meet latency and reliability access requirements, URLLC devices  access the network via 
 a two-step contention-free  access procedure,
    and mMTC devices  access the network utilizing a contention-based timing advance
   (TA)-aided access mechanism to meet the  massive access requirement.

     \item We propose an attention-based LSTM prediction model to predict the number of  active URLLC devices.
     As such, the BS can determine the parameters of muti-user detection  dynamically  based on the reliability requirement.
     This further ensures all  active URLLC devices  to successfully access the network  in one shot to meet the latency requirement.
         %In addition, to reduce the overhead,  we predict the maximum number of new arriving active URLLC devices 
         % during several random access slots, and thus

    \item We present the numerical results, which show  that,
     compared to the benchmark schemes, the  proposed scheme significantly improves the successful access probability,
     and satisfies the  diverse QoS requirements of URLLC and mMTC devices. 
\end{itemize}

%The rest of this paper is organized as follows.
%In Sections II and III, we elaborate the Cellular Network Architecture and the proposed  LSTMH-RA scheme, respectively. In Section IV, we analyze the LSTMH-RA scheme. Sections V and VI give numerical results and the conclusion, respectively.
%\begin{figure*}[htbp]
%	\centering
%	\includegraphics[scale =0.65] {figures/Network_1.eps}\\
%	\caption{The network architecture of a heterogeneous MTC network for 5G-enabled smart city, including the
% terminals layer, networks layer, platforms layer and applications layer. }\label{system}
%\end{figure*}

\section{MTC Network Architecture based on 5G}
%The architecture of  IoT networks for smart city

%The 5G communication networks have become the main force supporting and promoting IoT services.
Due to the extensive coverage, security, reliability and
flexibility of communication of 5G
communication network, 5G network has become
the main force to support and promote MTC services.
In this section, we introduce the  MTC network architecture based on 5G, which includes four tiers~\cite{IOTSC}, including
 terminals layer, networks layer, platforms layer and applications layer.
 %, as shown in Fig.~\ref{system}.

\textbf{Terminals layer} includes MTC devices with access ability, which can be divided into two categories: mMTC devices
and URLLC devices. mMTC devices mainly include various sensors, such as  temperature sensors and  humidity sensors,  and URLLC devices include
medical equipments and industrial equipments, etc.

\textbf{Networks layer} is the 5G
cellular network and other access networks interconnected with the 5G network. With powerful cloud technologies, some  functions originally performed in the  base station (BS)
and core networks, including access control, power control,  signal processing,  and so on, are migrated to the
cloud~\cite{cloud}.

\textbf{Platforms layer} refers to the IoT service platform, including  cellular network operator platforms, enterprise  platforms,
government platforms, etc.

\textbf{Applications layer} covers smart industrial,  smart manufacturing, autonomous driving,  and other services.

\section{The proposed IHRA scheme for 5G-enabled smart city}
In this section, we first introduce the traditional contention-based random access scheme in the existing cellular network,
 and  then describe our proposed IHRA scheme for 5G-enabled smart city.

\subsection*{A. Traditional contention-based random access scheme}

In the Long Term Evolution-Advanced \mbox{(LTE-A)}  cellular network, according to different service triggering events,
 random access schemes can be divided
 into two categories: contention-based random access procedure and non-contention based random access procedure.  The
contention-based random access procedure makes users to select their preambles in a contention manner, and 
the non-contention based random access procedure is to pre-allocate preambles to some users before random access procedure.
Due to the sporadic transmission of devices in MTC communications, the non-contention based random access procedure is infeasible. In the following,
we briefly introduce the traditional contention-based random access process.

%there are two kinds of random access schemes: one is

\begin{figure*}[htbp]
	\centering
	\includegraphics[scale =0.65] {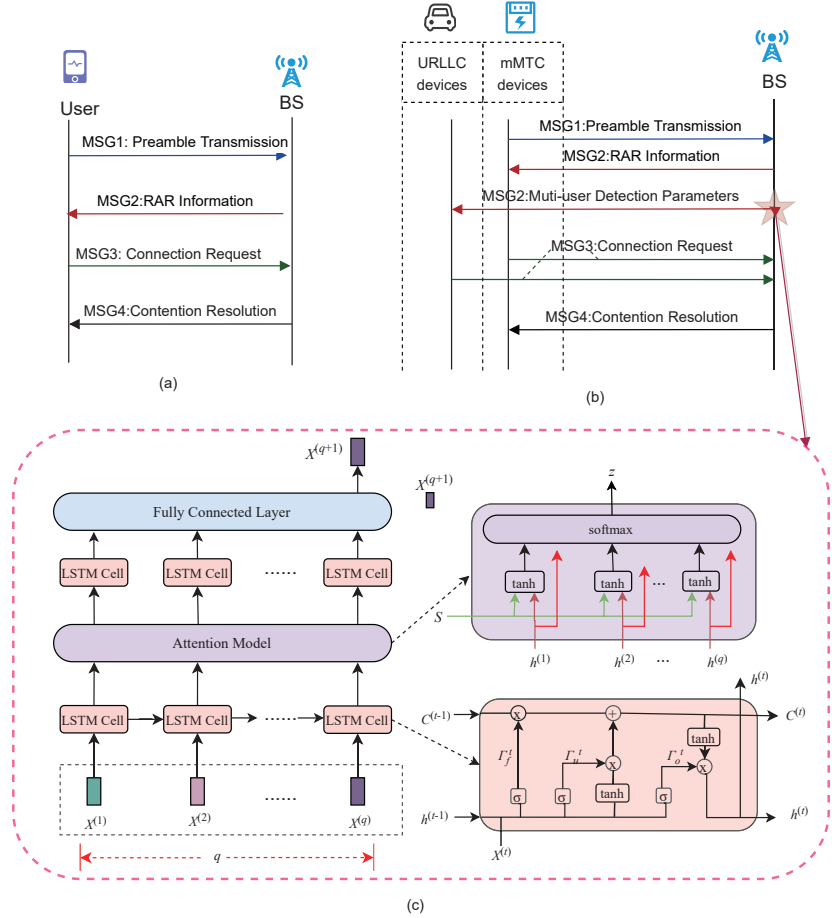}\\
	\caption{Random access schemes: (a) the traditional contention-based random access scheme; (b) the proposed random access procedure
 including (c) the proposed attention-based LSTM prediction model.}\label{PR}
\end{figure*}

As shown in Fig.~\ref{PR}(a), the contention-based random access process requires the interaction of four messages
(i.e., MSG 1, MSG 2, MSG 3, and MSG 4) between the BS
 and users, which is designed for H2H communications with large data packet transmission, few connections, and low energy consumption requirements.
  If these four messages can be successfully exchanged, an access request is finally completed. 
   The term ``contention'' means
that multiple users  send the same
 preamble sequence to the BS in the current random access slot to obtain the BS's resource grant,
  and the BS cannot figure out which users send this preamble. Hence, users need to send a
   unique message (MSG 3)  to the BS, and the BS will  transmit a message (MSG 4) to  users  to confirm which users access the network successfully.
  In addition,  the BS usually broadcasts a system information to inform the  available preamble set before the random access procedure. 
 The details are described as follows:

 \textbf{{MSG 1: Preamble Transmission}}

 In the current random access slot, each user randomly selects a preamble from the available preamble set,  and sends it to the BS via  physical random access channel (PRACH).
 Since the available preambles are orthogonal, if more than one  users select the same preamble in the current random access slot,
  the BS cannot distinguish them, resulting in the
  preamble collision.

\textbf{MSG 2: Random Access Response (RAR)}

The BS detects the received preamble signal.  Upon  preamble collisions, the BS
may fail to detect the transmitted preamble.  If a preamble is detected successfully, the BS sends a 
RAR message mainly  including the detected preamble
confirmation information and  the uplink resources used to transmit MSG 3.

\textbf{MSG 3: Connection Request}

If the user receives the RAR information corresponding to
its selected preamble within the waiting time, the connection
request message is transmitted on the allocated uplink resources.
If multiple users leverage the same uplink resource to transmit the connection  request message, collisions will occur.

 \textbf{MSG 4: Contention Resolution}

If the BS successfully receives the connection  request message,
 it sends a contention resolution message to the user,  indicating that the user has successfully accessed the network.
   After the user receives this information, the random access process ends.
Then, after the user and the BS  go through a series of
 higher-level signaling,  the user can transmit data information to the BS.

The aforementioned random access procedure is proposed for H2H communications.
There are several challenges when it is utilized for the MTC communications:
\begin{itemize}
\item Compared with the small data packet transmitted by the MTC devices,  the above random access procedure will introduce
heavy signaling overhead, thereby reducing the efficiency of data transmission.
\item Massive device accessing the network  will cause serious  preamble collision. This further increases the number of
retransmission devices, thereby increasing  delay and power consumption.
\item The above random access procedure does not take the different QoS requirements into consideration, which is
        not suitable for the heterogeneous MTC network where URLLC and mMTC devices \mbox{co-exist}.
\end{itemize}

The combination of deep learning and wireless communication is the key technology to realize the ``smart connection" in 5G-enabled smart city.
Therefore, to tackle the above problems, we propose an IHRA scheme  by utilizing LSTM and attention models.

%this project uses deep learning methods to propose a smart authorization-free
%
% random access solution suitable for future 6G cellular Internet of Things MTC communication,
% to take into account the communication requirements and ultra-wide coverage requirements of two different MTC application scenarios,
% mMTC and URLLC , Alleviate pilot frequency collision, improve the number of successfully connected devices and resource utilization,
% thereby reducing delay and power consumption. It aims to provide theoretical support and technical guidance for future 6G cellular
% Internet of Things related technologies, so as to promote the development of Internet of Things related industries and fully realize
%  the intelligent connection of all things, which has very important research significance.

\subsection*{B. The proposed IHRA scheme}
Fig.~\ref{PR}(b) illustrates the proposed IHRA scheme,
which aims to satisfy the diverse QoS requirements by utilizing  deep learning technology.
In addition, the cell is divided into multiple annulus with quantized distance $d=16T_s\times c$ from the center to the edge of the cell with radius $R$,
where $T_s$ is the basic time unit in the communication system and $c$ is the speed of light~\cite{zhang}. Devices in the same annulus
have the same TA index, and the number of annulus in the cell is $\zeta=\lceil{\frac{R}{2d}}\rceil$. We assume that each mMTC
device  knows its distance to the BS by
 utilizing distance measuring technologies,
and thus it knows its TA index. This assumption is reasonable since most mMTC devices' locations are fixed~\cite{zhang}. The details of the proposed IHRA scheme are described as follows.

%We use  $TA_{i}$ to denote the TA indexes of devices located at the $i^{th}$ annulus.
% For example, as shown in Fig.~{\ref{System}}, TA indexes of devices n1, n2, n3 and n4
%  are  $TA_{1}, TA_{1}, TA_{2}$, and $TA_{3}$, respectively.
% Furthermore,  each mMTC device can know its distance to the BS by utilizing
% distance measuring technologies, and thus it knows its TA information.
% In the proposed scheme, to meet the massive access requirement of mMTC devices,
% the proposed random access scheme utilizes the TA information to  distinguish devices selecting the same preamble.

%Specifically, to meet the latency and reliability requirements,
 %URLLC devices access the network via a contention-free access mechanism; to meet the massive access requirement,
 % mMTC devices  access the network via a  contention-based TA-aided access mechanism. In addition,
 % we propose an attention-based LSTM prediction model to predict the number of active URLLC devices,
 %  and thus the parameters of multi-user detection can be configured  dynamically. The details are described as follows.

\textbf{MSG 1: Preamble Transmission}

%Before  random access scheme, the BS predicts the number of active URLLC devices. Then, based on the predicted number of active URLLC devices, the BS selects the modulation and code scheme (MCS) which satisfies the reliability of URLLC devices transmission, and broadcasts the MCS to active URLLC devices. Since the number of URLLC devices is usually small, each device is allocated a unique spread code.
%Based on the broadcasted MCS, each URLLC device modulates and codes their payload data to the BS.

%The BS first predicts the number of URLLC devices to be accessed in next RA slot by utilizing an attention-based LSTM prediction model and dynamically reserves preamble resource.

 %Each URLLC device transmits its dedicated preamble directly to  the BS. Assume that ${\tau_|text{URLLC}}$ preambles are reserved for URLLC %devices.

Each active mMTC device randomly selects a preamble sequence from the available $\tau_p$ preambles, and mMTC device in the $i^{th}$ annulus
regards the $[\frac{26-\zeta}{2}+i]^{th}$
 subcarrier as the starting position to place their selected preambles~\cite{zhang}.
 %broadcasted by the base station and transmits it in the next available RA slot.In addition,each mMTC device also  randomly selects a power level from others except the highest power level broadcasted on Physical Broadcast Channel.
% To alleviate preamble collision, we adopt the preamble placement method proposed in~\cite{zhang}~to make the BS estimate the number
% of devices selecting the same preamble and having  the same TA index (i.e., in the same annulus).
% Specifically, the mMTC device in the $i^{th}$ annulus regards the $[\frac{26-\zeta}{2}+i]^{th}$
% subcarrier as the starting position to place their selected preambles.
 %By doing so, the central ${838+\zeta}$ subcarriers should be used to carry these preambles and the rest subcarriers for grand band[12].

 %Specifically,We first divide the cell into ${\zeta}$ annuluses with the quantized 16${T_s}{c}$ .

Let $\bm{\rho}_{r,i}$ denote the received  preamble $r$ from the $i^{th}$ annulus,
and $n(r,i)$ denote the number of devices selecting preamble $r$ in  the $i^{th}$ annulus.
 Then, the received preamble signal  can be written as $\bm{Y}=\sum_{r=1}^{\tau_p}\sum_{i=1}^{\zeta}{n(r,i)}\bm{\rho}_{r,i}+\bm{N},$
% \begin{equation}\label{1}
% \begin{array}{l}
%  \bm{Y}=\sum\limits_{r=1}^{\tau_p}\sum\limits_{i=1}^{\zeta}{n(r,i)}\bm{\rho}_{r,i}+\bm{N},
% \end{array}
% \end{equation}
 where $\bm{N}$  represents additive white Gaussian noise with mean zero and variance $\sigma^2$.

\textbf{MSG 2: RAR  Transmission and Muti-User Detection Parameters Broadcasting}

%The eNB first detects the received preamble.If the eNB receives the preamble that is reserved for a specific URLLC device, it further retrieves the information of the device and then allocates resource blocks for the device based on the addresses assigned during the initial registration process.Otherwise,

The BS can obtain  the value of $n(r,i), (r=1,\dots,\tau_p, i=1,\dots,\zeta)$,  
based on the cross-correlation value of $\bm{Y}$ and preamble $r$~\cite{zhang},
which is discussed in~\cite{zhang} in detail.
 %The details procedure on the estimation of $n(r,i)$ and the TA information selection is given in Algorithm 1.
 Then, the BS generates a RAR  corresponding to  preamble $r$ with $n(r,i)=1$, including preamble identification, TA index, and  resource block.
 Note that, $n(r,i)=1$ means that only one device selects preamble $r$ in the $i^{th}$ annulus. Thus,
 there are multiple RARs corresponding to preamble $r$, resulting in preamble $r$ corresponding to multiple resource blocks.

%\begin{algorithm}
%	\SetKwInOut{KIN}{Initialize}
%	\SetKwInOut{Return}{Return}
%	\caption{The method of $n(r,i)$ estimation}
%\KIN {$n(r,i)=0$}
%
%
%\For{   $r\in \{ 1,2,\dcots,\tau_P\}$ }
%{
%	\For{   $i\in \{ 1,2,...,\zeta\} $ }
%	{
%		\If{$\left|\sum\limits_{r=1}^{\tau_p}\bm{Y}\cdot\rho_{r,i}(t)\right| >\gamma$}
%		{
%			$n(r,i)++$\;
%			$\bm{Y}=\bm{Y}-\rho_{r,i}$\;
%			
%		}
%		
%	}
%%\eIf{$n(r,i)==1$}
%%{
%%	$T_A$=$T_{A,i}$\;
%%	eNB send RAR\;
%%}{
%%eNB stop sending RAR\;
%%}
%}	
%\end{algorithm}

 %In addition,  we  predict the maximum  number of active URLLC devices during  consecutive $L$ random access slots instead of every random  access slot to reduce the overhead.   If the current random access slot is the prediction slot,
% In addition, the BS predicts the predict the maximum  number of active URLLC devices during  consecutive $L$ random access slots  by utilizing our proposed attention-based LSTM prediction model. Otherwise, the BS can directly obtains the  number of active URLLC devices based on the previous prediction.

In addition, by utilizing our proposed  attention-based LSTM prediction  model,  the BS predicts the number of active URLLC devices.
 Thus,  based on the reliable transmission requirement, the BS determines  and broadcasts the parameters of muti-user detection
  (including the resource block, modulation and code schemes) to all active URLLC devices.

% selects the modulation and code scheme (MCS)  satisfying the reliability of URLLC devices transmission,  allocates the corresponding resource blocks, and broadcasts the MCS and resource blocks to active URLLC devices.

 %Since the number of URLLC devices is usually small, each device is allocated a unique spread code.
%Based on the broadcasted MCS, each URLLC device modulates and codes their payload data to the BS.

\textbf{MSG 3:  Uplink Message Transmission}

%Each URLLC device transmits data information based on the resource block allocated.
 Each mMTC device  finds  RARs including its preamble identification, and  matches the TA indexes in these RARs with its own.
 If the TA index of one RAR is the same as its own, it  transmits its uplink message with the highest transmit power level via the
 resource block indicated by this RAR. Otherwise, this mMTC device randomly selects a resource block from resource
 blocks allocated to its selected preamble, and  transmits its uplink message with a randomly selected transmit power level.

 Based on the broadcasted parameters,   active URLLC devices modulate and code their payload data to obtain their uplink messages,
 and transmit  to the BS via the same resource block.
 %to transmit data information from the data transmission resources allocated to the preamble selected by the device. If there is no data transmission resources under the preamble identity field by the devices, the mMTC devices end up transmitting preambles repeatedly until the maximum allowed number of preamble transmissions is reached. Then the mMTC devices declare access failure and exit the RA procedure.

\textbf{MSG 4: Contention Resolution}
%Since the eNB dispatches reserved preambles for each URLLC device, it ensures that the access process of URLLC devices is collision-free. Furthermore it allocates resource blocks based on the addresses of URLLC devices that assigned during the initial process, which guarantees the successful transmission of data information. So it's no need to have a contention resolution process for the URLLC devices.

For each resource block  carrying the uplink message of mMTC devices, the BS utilizes the successive interference canceller (SIC) algorithm
 to detect the received uplink message~\cite{2015Consideration}.  The SIC algorithm decodes the 
 uplink message of devices from the highest to the lowest power level. More specifically, the mMTC devices with 
 the highest power level will be decoded first. If it can be successfully detected, 
 the interference of the data information of this device is eliminated, 
 and then the data information of the device with the second highest power level 
 is detected until the data information of the device cannot be successfully detected. 
 %It should be noted that if the number of devices selected the $l^{th}$ power level
% on the same resource block is 1, (provided that the data information of the devices
% with the $1st$ to $l-1th$ power levels has been successfully detected), the data
% information of this device can be successfully detected. Otherwise,
% the BS cannot successfully detect the data information of the remaining devices
% in this resource block, and thus this device failed to access the network.
%More specifically, mMTC devices with the highest power level are decoded first.
The following two events ensures that  device selecting power level $l$ can be successfully decoded:
1) this device is free from power level collision  and  devices with power levels  larger than  $l$
 are successfully decoded; 2) this device is free from power collision  and  the number of devices
 with power levels  larger than  $l$ is zero. If this device can be decoded successfully,
 the interference of the uplink message of this device is cancelled from the received uplink message.
 %Otherwise, the device fails to access the network and will access the network in the upcoming random access slot.

 The BS decodes the uplink messages of URLLC devices by utilizing the muti-user detection algorithm,
 which is not the focus of our paper and interested readers can refer to~\cite{mul1} and references therein.

\subsection{Attention-based LSTM  prediction model}

The  proposed random access scheme predicts the number of active URLLC devices by utilizing
a proposed attention-based LSTM  prediction model, and configures the parameters of  multi-user detection to these URLLC devices,
to guarantee the reliability and latency requirements  of URLLC devices.  The proposed prediction model includes  two LSTM layers,
 one attention layer and one fully connected layer, as shown in Fig.~\ref{PR}(c).
 The  data set travels from the  first LSTM layer to the attention layer,  then feeds into the second LSTM layer,
  and finally connects to a fully connected layer to predict the number of active URLLC devices.
  We describe the details of each layer of this prediction model as follows.

%\begin{figure*}[htbp]
%	\centering
%	\includegraphics[scale =0.5] {figures/LSTM.eps}\\
%	\caption{ Attention-based LSTM prediction model consisting of  two LSTM layers, one attention layer and one fully connected layer.}\label{LSTM-model}
%\end{figure*}

\subsubsection{LSTM layer} 
%LSTM was first proposed by Hochreiter & Schmidhuber in 1997. The original intention of the design was to 
 
\

 LSTM is a special kind of recurrent neural network (RNN)~\cite{LSTM}. In the original RNN, as the  increase of training time  and the
 number of network layers, the problem of gradient explosion or gradient disappearance is prone to occur, which makes
 it impossible to process long sequence data, and thus cannot obtain information about long-distance data. LSTM is  
 designed to let neural networks remember  long-term information.
The application areas of LSTM  mainly includes text generation, machine translation, speech recognition, image description generation 
and video tagging, etc.

 Each LSTM layer consists of multiple LSTM cell.
Each LSTM cell includes  input gate,  output gate, and forget gate.
The input data of the LSTM is denoted by $\bm{X}=[X^{(1)},X^{(2)},\cdots,X^{(q)}]$,
where $q$ is the length of the time steps. The process of building these gates  during the $t^{th}$ time step is~\cite{2016LSTM}
\begin{equation}\label{lstm}
\begin{array}{l}
\left\{
\begin{array}{l}
\bm{\Gamma_{f}^{(t)}}=\delta(\bm{W^{(f)}[h^{(t-1)},X^{(t)}]+b^{(f)})} \\
\bm{\Gamma_{u}^{(t)}}=\delta(\bm{W^{(u)}[h^{(t-1)},X^{(t)}]+b^{(u)})}\\
\bm{{\tilde{C}}^{(t)}}=\tanh(\bm{W^{(c)}[h^{(t-1)},X^{(t)}]+(b^{(c)})}\\
\bm{{C}^{(t)}}=\bm{\Gamma_{f}^{t}\times C^{(t-1)}+\Gamma_{u}^{(t)}\times \tilde{C}^{(t)}}\\
\bm{\Gamma_{o}^{(t)}}=\delta(\bm{W^{(o)}[h^{(t-1)},X^{(t)}]+b^{(o)})}\\
\bm{h^{(t)}}=\bm{\Gamma_{o}^{(t)}}\times \tanh(\bm{C^{(t)})},
\end{array}
\right.
\end{array}
\end{equation}
where $\bm{\Gamma_{f}^{(t)}}$, $\bm{{C}^{(t)}}$ and $\bm{h^{(t)}}$
are the forget gate, the input gate,
and the output gate, respectively,
$\bm{W}$ and $\bm{b}$ represent the weight
and the bias,
respectively. Furthermore, $\delta$  and $\tanh$ stand for  $sigmoid$ activation function and hyperbolic tangent function, respectively.
 %and $h^{(t)}$ is the output of  a LSTM cell.

\subsubsection{Attention layer}
\

%Attention  pays dynamic attention on some
%useful input information, which has been utilized to improve the performance of LSTM network~\cite{2014Neural}.

Attention model (AM) was first introduced from the machine translation task and has now become a mainstream neural
 network concept~\cite{2014Neural}. Attention can integrate related information and allow the model to provide dynamic attention to some useful input information,
 which has been utilized as a useful tool for improving the performance of LSTM network.

%which improves the performance of the model.In this subsection,we introduce the attention mechanism to improve the weight of the peak value in URLLC device prediction. The mechanism is designed to take advantage of any correlation between input features and output values.

%\begin{figure*}[htbp]
%	\centering
%	\includegraphics[scale=0.8]{figures/LSTMp.eps}\\
%	\caption{Prediction results: (a)  the prediction performance of our  proposed attention-based LSTM model,
%% and (b) comparison between the % proposed attention-based LSTM %and LSTM prediction models.}\label{LSTMP}
%\end{figure*}

\begin{figure}[htbp]
	\centering
	\includegraphics[scale=0.5]{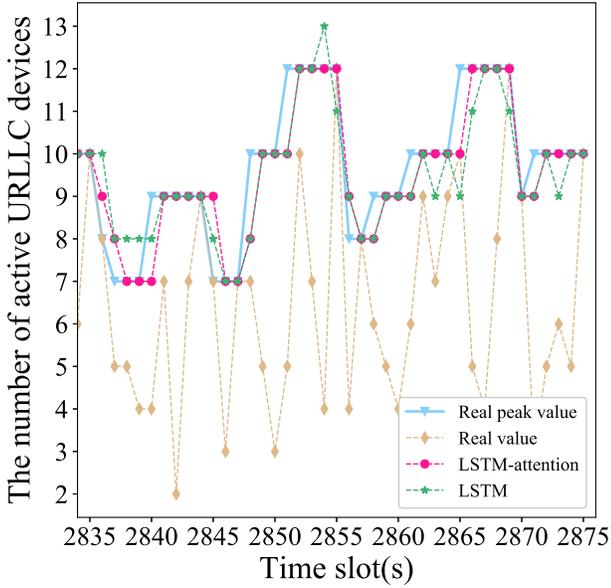}\\
	\caption{ The comparison between the  proposed attention-based LSTM and LSTM prediction models.}\label{LSTMP}
\end{figure}

%\begin{figure*}
%\renewcommand{\captionlabelfont}{\small}
 % \centering
 % \subfigure[]{
   % \label{fig:subfig:a} %% label for first subfigure
   % \includegraphics[scale=0.4]{figures/LSTMa.eps}}
 % \vspace{1in}
 % \subfigure[]{
   % \label{fig:subfig:b} %% label for second subfigure
   % \includegraphics[scale=0.4]{figures/LSTMb1.eps}}
     %\hspace{1in}
 % \caption{Prediction results: (a)  the prediction performance of our  proposed attention-based LSTM model, and (b) comparison between the  proposed attention-based LSTM and LSTM prediction models.}\label{LSTMP} %% label for entire figure
%\end{figure*}

%which improves the performance of the model.In this subsection,we introduce the attention mechanism to improve the weight of the peak value in URLLC device prediction. The mechanism is designed to take advantage of any correlation between input features and output values.

%Assume the input of the $k^{th}$ LSTM layer is $X = ({x_{k}^1},{x_{k}^2},...,{x_{k}^{q}})$ where ${q}$ is the length of time steps.
 Denote the output of the $t^{th}$ time step
 of the first layer by $\bm{h_{1}^{(t)}} = f(\bm{x_{1}^{(t)}},\bm{h_{1}^{(t - 1)}},\bm{C_{1}^{(t - 1)}})$
 where $f( \cdot )$ stands for  the LSTM process calculated by~(\ref{lstm}). Then,
the output of the attention layer during the $i^{th}$  time step  is $\bm{z^{(i)}} = \sum\limits_i {a^{(i)}\bm{h_1^{(i)}}},$ 
% \begin{equation}\label{Ag}
% \bm{z^{(i)}} = \sum\limits_i {a^{(i)}\bm{h_1^{(i)}}},
% \end{equation}
 where $a^{(i)}$ is the attention weight, which is computed by $a^{(i)} = \frac{{\exp (e^{(i)})}}{{\sum\nolimits_{j = 1}^{q} {\exp (e^{(j)})} }},$ 
%  \begin{equation}\label{Ag1}
% a^{(i)} = \frac{{\exp (e^{(i)})}}{{\sum\nolimits_{j = 1}^{q} {\exp (e^{(j)})} }},
% \end{equation}
 where $\exp$  stands for the exponential function, and  $e^{(i)}$ is  the aggregation state computed via the Bahdanau-attention method~\cite{2014Neural}: $e^{(i)}= \bm{V^T} \tanh (\bm{{W_u}s + {W_h}h_1^{(i - 1)}}),$ 
% \begin{equation}\label{Ag2}
% e^{(i)}= \bm{V^T} \tanh (\bm{{W_u}s + {W_h}h_1^{(i - 1)}}),
% \end{equation}
 where $\bm{V}, \bm{W_u}$ and $\bm{W_h}$ are the input weights, and
$\bm{s}=[\bm{{h_{1}^{(1)}},\cdots, {h_{1}^{(q)}}}]$ is  the output of LSTM during these  ${q}$  time steps.

Finally, we  combine  the attention model output $\bm{z^{(i)}}$
 and the LSTM output $\bm{h_1^{(i - 1)}}$, and regards  the combination signal as the input data for the current time slot,
  i.e. $\bm{X^{(i)}_1}=concat(\bm{z^{(i)}},\bm{h_1^{(i - 1)}})$ where   $concat$ is the concatenating function.

\subsubsection{Output layer}
\

 The extracted features in the last LSTM layer connect to
 a fully connected layer to obtain the output  $X^{(q+1)}$.
 After rounding the output  $X^{(q+1)}$, we get the predicted number of active URLLC devices.
 % Note that, the output $X^{(q+1)}$ may not an integer, and we should  round $X^{(q+1)}$ after prediction.

\section{Numerical Results}

We first present the prediction performance of our attention-based LSTM prediction model.
Then,   we make a comparison  between our proposed  IHRA scheme and the traditional  contention-based random access scheme (termed `TARA' in
our result figures), in  terms of  the number of
successful access devices. Note that, our proposed IHRA  scheme allocates the highest transmit power level  to devices  whose TA indexes
are the same as those in RARs corresponding to their selected preambles. To show the advantage of this allocation strategy, we also include a  scheme where
 all devices  select their uplink transmit powers randomly and uniformly  as our baseline. This scheme is termed `IHRA-random' in our result figures.
Furthermore, we set the quantized unit to $2d=32T_sc=156m$ where $T_s=3.072\times 10^{-7}s$  and  $c=3\times10^8m/s$ stand for the basic time
unit and the speed of light, respectively~\cite{zhang}.  For  our proposed attention-based LSTM prediction model, the learning rate is 0.001,
 and the loss function is the root-mean square function. The number of active URLLC devices in each time slot follows a distribution of
  Poisson with mean $\lambda=5$.

Note that, if the predicted number of active URLLC devices  is lower than the real one, our proposed IHRA scheme cannot ensure
 all active URLLC devices  access the network successfully in one shot. To tackle this problem, we take the maximum  number of active URLLC devices from
 $t$ to $t+4$ time slots  as the  value in the $t^{th}$
time slot.  Fig.~\ref{LSTMP}compares the  proposed attention-based LSTM and LSTM prediction models.
%The LSTM prediction model is obtained by deleting the attention model from our attention-based LSTM model.
In this figure, `real peak value'  stands for the maximum number of active URLLC devices from $t$ to $t+4$ time slots,
 and `real value' denotes the actual number of active URLLC devices in the $t^{th}$ time slot.
Fig.~\ref{LSTMP}(b) shows that our proposed attention-based LSTM prediction model predicts more accurately than the LSTM prediction model, and
 the predicted value is  larger than or equal to the real value in the $t^{th}$ time slot.
 This means that the  allocated  parameters can ensure  reliable
 communication for URLLC devices, so that  all active  URLLC devices during time slot $t$ can access  the network successfully.

%The comparison results consisting of the real and prediction traffic flows are shown in Fig.5. Although there is insufficient historical training data, it is clear that the trend of the peaks and valleys of the predicted URLLC devices is almost identical to the actual one. Fig.6 compares the proposed attention-based LSTM and basic LSTM prediction algorithms.The experimental results show that our algorithm can produce better prediction accuracy.

\begin{figure}[htbp]
	\centering
	\includegraphics[scale=0.5]{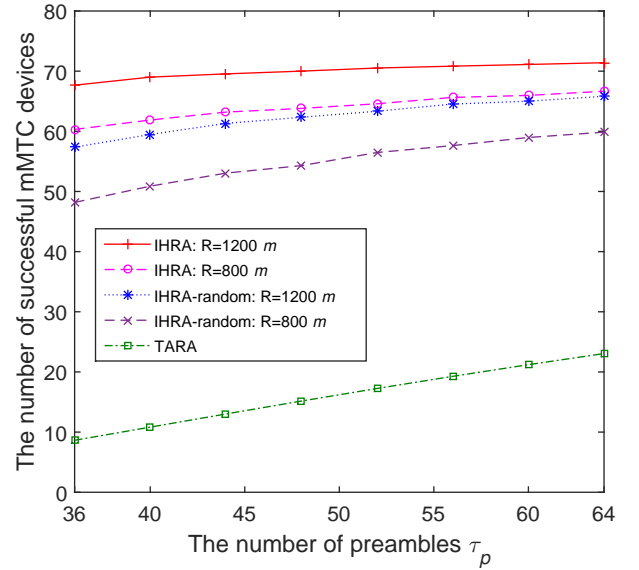}\\
	\caption{ The number of successful mMTC devices versus the number of preambles.}\label{TMtaop}
\end{figure}

Fig.~\ref{TMtaop} shows how the number of successful mMTC devices varies
with the number of preambles for $R=1200,800$.
 We set the number of active mMTC devices to $N_a=80$, and the number of power levels to $L=4$.
 We can see from Fig.~\ref{TMtaop}  that, with the increase of the number of preambles,
the number of successful mMTC devices increases and is significantly higher than the baselines. This indicates that the transmit power level allocation
 strategy is efficient to improve the number of successful devices. Furthermore, with the increase of the size of the  cell, the
 number of successful mMTC devices increases. The reason is that, with the increase of the size of the cell,
   the number of different TA values increases,  and then the BS can distinguish more mMTC devices.

%\begin{figure}[htbp]
%	\centering
%	\includegraphics[scale=0.45]{figures/T-L-N.eps}\\
%	\caption{The number of successful mMTC devices versus the number of power levels.}\label{TML}
%\end{figure}
%
%
%
%Fig.~\ref{TML} shows how the number of successful mMTC devices
%changes with the number of power levels for $R=1200,800$ and $400$.
%We set the number of active mMTC devices to 80 and the number of preambles to $64$.
%We can see from Fig.~\ref{TML}  that, with the increase of the number  of power levels,
%the number of successful mMTC devices increases and is significantly higher than the baselines. Furthermore, with the
%increase of the size of the  cell, the number of successful mMTC devices increases. The reason is the same as we described in Fig.~\ref{TMtaop}.
\begin{figure}[htbp]
	\centering
	\includegraphics[scale=0.5]{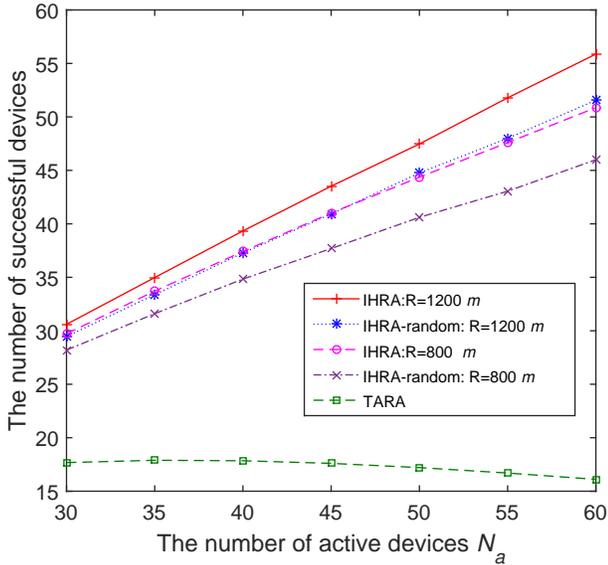}\\
	\caption{ The number of successful  devices versus the number of active devices.}\label{TN}
\end{figure}

Fig.~\ref{TN} shows how the number of successful  devices changes
with the number of active devices for $R=1200,800$.
 We set  the number of preambles to 40, the number of active URLLC devices to
 3. We can see from Fig.~\ref{TN}  that,
 with the increase of the number of active devices, the number of  successful devices decreases slowly for the TARA scheme  and increase linearly  for 
 other schemes. We can also note that 
  the number of successful devices of our proposed IHRA scheme is significantly higher than the baselines.
  Furthermore, with the increase of the size of the  cell,  the number of successful mMTC devices increases.
   The reason is the same as we described for Fig.~\ref{TMtaop}.

\section{Conclusion  and future directions}
IoT is the enabler for smart city  to interconnect devices, which makes MTC account for a great proportion in all communication
services.  mMTC and URLLC will co-exit in MTC network. mMTC requires
  massive connections, while URLLC requires high reliability and low latency communications. In this article,
 we introduce the  architecture of heterogeneous MTC network, and then propose an IHRA scheme
  for 5G-enabled smart city.
%An attention-based  LSTM prediction \mbox{model} is introduced to predict the number of active URLLC devices.
 Numerical results show  that, compared to the  benchmark schemes, the proposed scheme significantly improves the successful access probability, 
 and satisfies the  diverse QoS requirements of URLLC and mMTC devices. 

\bibliographystyle{IEEEtran}
\bibliography{ref}
\end{document}